\documentstyle[preprint,aps]{revtex}
\begin{document}
\draft
\title{A Soluble Free-Fermion Model in $d$ Dimensions}
\author{F. Y. Wu and H. Y. Huang}
\address{Department of Physics, Northeastern University, Boston,
	 Massachusetts 02115}
\date{\today}
\maketitle

\begin{abstract}
We consider a vertex model in $d$ dimensions characterized by
lines which run in a preferred direction.
We show that this vertex model is soluble if
the  weights of vertices with intersecting lines
are given by a free-fermion condition, and that a fugacity $-1$
is associated to each loop of lines.
The solution is obtained by mapping
the model into a dimer problem and by
evaluating a Pfaffian.
We also determine the critical point and the singular behavior of the
free energy.
\end{abstract}
\vskip 1cm
\section{INTRODUCTION}

An outstanding unsolved problem in statistical mechanics has been the
exact solution of lattice models in three or higher dimensions.
While a host of lattice models has been solved in two dimensions \cite{rf1},
very few exact results are known for three or higher dimensions.
Some two decades ago Suzuki \cite{rf2}  solved a three-dimensional Ising model
with pure 4-spin interactions, which turns out to be a two-dimensional
model in disguise.
Similarly, a three-dimensional 5-edge model solved by Orland \cite{rf3}
is equivalent to a two-dimensional dimer problem \cite{rf4}.
The first exact solution of a  genuine   three-dimensional model
was that obtained by Baxter  \cite{rf5} for
the Zamolodchikov model \cite{rf6},   a 2-state
interaction-round-a-cube (IRC) model involving some negative Boltzmann
weights.  Since then
Baxter and Bazhanov \cite{rf7} have further
extended the solution to the IRC model of
$N$ states for general $N$.
Very recently, we have solved
a vertex model in arbitrary dimensionality \cite{rf8}.
This is a vertex model described by
lines which run in a preferred direction on the lattice
and do not intersect.
Like the three-dimensional IRC model of Refs. \cite{rf5} and \cite{rf6},
this model also involves negative Boltzmann weights.
While the nonintersecting feature  makes it possible to use this model to
describe  flux lines in type II superconductors \cite{rf9}, the more general
version
of the  model with vertices of intersecting lines remains unsolved.

Some twenty-five  years ago one of us and Fan \cite{rf10,rf11}
introduced the term
{\it free-fermion model} to lattice-statistical vertex models
satisfying a certain free fermion condition.
When this condition holds, the model can be regarded as
one describing  free fermions on a lattice \cite{rf8} and becomes
soluble.
But soluble free-fermion models have
been restricted mainly to two dimensions \cite{rf12}.  In
this paper we extend the solution of \cite{rf8} to a free fermion  model
which includes vertices of intersecting lines in
any dimension.
This yields a new class of
soluble free fermion models in arbitrary dimensionality.

\noindent
\section{THE MODEL}

In the vertex models considered in \cite{rf8}, vertex configuration are
described by lines embedded on a lattice subject to:

i) lines do not intersect,

ii) lines are oriented in a preferred
direction, and

iii) each loop  of line possesses
a fugacity $-1$.

Now we relax condition (i) by allowing lines to intersect.
For definiteness we consider a Cartesian lattice ${\cal L}$ in $d$
dimensions.   Place bonds along edges of ${\cal L}$ with the restriction
that the number of bonds incident at
each  vertex
is zero or even, with half of the bonds
incident in the positive and the other half in the negative axes directions.
The bonds then form lines which  run along the main diagonal
of ${\cal L}$.
Thus, we have a $D$-vertex model with $D$
different vertex configurations,
where
\begin{equation}
D= \sum_{r=0}^{d}
 {d\choose{r}}^2 ={{2d}\choose{d}},
\end{equation}
where $r$ is the number of lines intersecting at the vertex.
This leads to $D=6,20,....$ for $d=2,3,...$, respectively.

Number the
negative axes  incident at a vertex
$\{ n\} =\{1,2, ..., d\}$ and the
positive incident axes $\{p\}=\{d+1, d+2,...,2d\}$,
such that $\{i,d+i\}$ refer
to the axis in the $i$th direction.
Denote by $\omega_{\{n'\} \{p'\}}$ the weight of a vertex
with bonds (lines) incident at
$\{n'\}\in\{n\}$ and $\{p'\}\in\{p\}$.
For vertices with no incident bonds we take the weight
$\omega_0=1$, and for vertices with
two incident bonds, one in the negative direction
$i$ and one in the positive direction $j$, we take the weight
to be $\omega_{ij}$.  For
vertices  with four or more bonds
the weights are  given by the free-fermion condition
\begin{eqnarray}
\omega_{\{n'\} \{p'\}} & = & \sum_P(-1)^{\delta_P}
			 \prod_{i\in \{n'\},j\in \{p'\}}
			  \omega_{ij} \nonumber \\
		       & = & {\rm det} |\omega_{ij}|,
\end{eqnarray}
where the summation is taken over all pairings $P$ of
axes into pairs $\{ij\}$,  $i\in\{n'\}$ and $j\in\{p'\}$,
and $\delta_P$ is the signature of $P$,
the number of
transposition needed to bring the indices of the  pairing to the
{\it canonical ordering}
\begin{equation}
\{1,d+1\}\{2,d+2\}\cdots\{d,2d\}.
\end{equation}
Here, it is understood that indices not contained
in $\{n'\}$ and $\{p'\}$ are deleted in (3) in determining
the canonical ordering.
For $d=3$, for example,
(2) gives
\begin{mathletters}
\begin{eqnarray}
\omega_{1245}  & = & \omega_{14}\omega_{25} - \omega_{15}\omega_{24} \nonumber
\\
\omega_{1256}  & = & \omega_{16}\omega_{25} - \omega_{15}\omega_{26}
\nonumber\\
\omega_{1246}  & = & \omega_{14}\omega_{26} - \omega_{16}\omega_{24},
\end{eqnarray}
\text{ for $\{n'\}=\{12\}$,
       and other similar relations for $\{n'\} =\{13\}, \{23\}$, and }
\begin{eqnarray}
\omega_{123456}
     & = & \omega_{14}\omega_{25}\omega_{36} +
\omega_{15}\omega_{26}\omega_{34}
	 + \omega_{16}\omega_{24}\omega_{35} \nonumber \\
     &   & \mbox{}  - \omega_{14}\omega_{26}\omega_{35}
	 - \omega_{15}\omega_{24}\omega_{36} - \omega_{16}\omega_{25}\omega_{34} ,
\end{eqnarray}
\end{mathletters}
for $\{n'\}=\{n\}=\{123\}$.
Note that the expression (2) can be written as a determinant as indicated.
The model considered in \cite{rf8} corresponds to taking
$\omega_{ij} =\sqrt {z_iz_j}, \>z_i=z_{i+d}$ so that
the weights (4a) and (4b)  vanish identically, indicating  that
vertices
with 4 or more bonds are forbidden.

The vertex weight (2) can be represented graphically
by decomposing the vertex configuration
into continuous lines  indicated by the pairings. That is,
for each $\omega_{ij}$ one connects the
negative axis $i$ to the positive axis $j$.
For the vertex weight (4b), for example, this leads to the graphs
shown in Fig. 1.
In this picture and assuming periodic
boundary conditions, lines
form loops wound around the lattice.
Let $\ell$ be the overall number of loops when each vertex configuration
is decomposed into its canonical ordering.
This permits us to define a partition function
\begin{equation}
Z = \sum _{\rm vertex\>config.}(-1)^\ell\prod_{\rm vertex}
  \omega_{\{n'\} \{p'\}}.
\end{equation}
Now, substitute (2) into (5) and observe
that
when all pairings are in the
order of the canonical form we have $\delta_P=0$,
and whenever the pairing of two lines
interchanges at a vertex, such as changing $\{14,25\}$ to $\{15, 24\}$
in the first line of (4a), both $\ell$ and $\delta_P$ change by 1.
It follows that
we can rewrite the partition
function (5) as
\begin{equation}
Z =  \sum _{\rm loop\>config.}(-1)^\ell\prod \omega_{ij},
\end{equation}
where the summation is taken over all possible loop configurations
of $\ell$ loops, and the product is taken over all $\omega_{ij}$ factors
only, which are present along
the loops.
In the next section we show that, in the form of (6),
the partition function $Z$
is precisely a dimer generating function.

\section{A DIMER PROBLEM}

Introduce a dimer lattice ${\cal L}'$
by expanding every lattice point of ${\cal L}$ into
a ``city", which we choose to be the one consisting of $2d$ points
located on the $2d$ incident axes $\{n\}$ and $\{p\}$.
The situation of $d=3$ is shown in Fig. 2.
Let $N$ and $2dN$ be the respect numbers of lattice points
in ${\cal L}$ and ${\cal L}'$.
Connect as shown in Fig. 2 each point in $\{n\}$
to every point in $\{p\}$, and vice versa.
There are now two kinds of edges in ${\cal L}'$
: intercity edges connecting cities
and intracity edges within a city.
Starting from a given vertex configuration on ${\cal L}$,
we cover each intercity edge on ${\cal L}'$ by a dimer
(resp. leave it empty) if the corresponding edge on ${\cal L}$
is empty (resp. covered by a bond).
The remaining uncovered points on ${\cal L}'$
are then covered by dimers placed along intracity edges.
This latter covering
corresponds to the different pairings of the remaining points and is
generally not unique.  For example,
the different coverings corresponding to
$\omega_{1245}$ and $\omega_{123456}$ given by (4a) and (4b)
together with the associated signs are shown in Fig. 3.
In this way, we have mapped  vertex configurations
into dimer coverings.

We now establish that there is a one-one correspondence between
dimer coverings of ${\cal L}'$ and loop configurations in (6).
Consider first dimer coverings on ${\cal L}'$.
Superimpose any given dimer covering $C_i$ with a standard one, $C_0$,
in which all intercity edges are covered by dimers.
The superposition of two dimer coverings produces a graph of transition
cycles, or polygons \cite{rf13}.
In the present case, the transition cycles belong
to one of two kinds: double dimers placed on intercity edges, and
polygons wound around the lattice forming loops.
Thus, each dimer covering is mapped into a loop configuration.
  Conversely, to each loop configuration in (6), there exists
a unique dimer covering $C_i$ which, when superimposed upon $C_0$,
produces the loop configuration in question. This completes the proof.

It is well-known that dimer coverings are generated by a Pfaffian,
provided signs can be fixed correctly
  \cite{rf13}.
In the usual dimer problem all terms in the dimer generating
function have the same sign, and for this reason the generating function
can be written as a Pfaffian only for planar lattices.
In the present case, however,
the partition function (6)
is precisely a Pfaffian as we now see.

Number the $2dN$ lattice points of ${\cal L}'$ in the increasing order
along the main diagonal of ${\cal L}$, and
within a city
in the order as shown in Fig. 2.
Direct all intercity edges of ${\cal L}'$ in the positive direction
and all intracity edges of ${\cal L}'$ as shown in Fig. 2.
In this way all edges  are directed
in the positive directions.
Further, the factor $\prod \omega_{ij}$ in (6) is generated by
taking the dimer weight
 1 for intercity edges and weights $\omega_{ij}, i<j$,
for intracity edges.  To fix the sign associated with this product, we
consider a typical term in the Pfaffian corresponding to the
dimer configuration $C_i$.  The sign of this term relative to the term
corresponding to $C_0$ is the product of the signs of the transition cycles
produced by the superimposition of $C_i$ and $C_0$ \cite{rf13}.
The rule is that each
transition cycle carries a sign $(-1)^{n+1}$, where $n$ is the number
of arrows pointing in a given direction when the transition cycle is
traversed.  For transition cycles consisting of double dimers we have
$n=1$ so that the sign is always positive.  For transition
cycles consisting of loops
around the lattice we have, since all edges are
directed in the same direction, $n={\rm even}$, so that
the sign is always negative  for each loop.  Then the overall
sign of $C_i$ relative to $C_0$ is
$(-1)^\ell$.
This establishes that  the partition function (6) is precisely a Pfaffian.

\section{EVALUATION OF THE PARTITION FUNCTION}

We now evaluate the Pfaffian.

Let $N=N_1\times N_2\times\cdots\times N_d$, where $N_i$ be the
linear dimensions of ${\cal L}$.
Then the
Pfaffian is the square
root of the determinant of a $2dN\times 2dN$ antisymmetric matrix  ${\bf A}$
in the form of a direct product of $2d\times 2d$ matrices indexed
by
\begin{eqnarray}
&& {\bf A}(n_1,...,n_d|n_1,...,n_d)=\left( \begin{array}{cc}
   {\bf 0} & {\bf W}  \\ -
   {\bf W} & {\bf 0}  \end{array}  \right) \nonumber \\
&& {\bf A}(n_1,...,n_i,...,n_d|n_1,...,n_i+1,...,n_d) = \left(
   \begin{array}{cc}  {\bf 0} & {\bf U}_i \\
    {\bf 0} & {\bf 0}  \end{array} \right)  \nonumber \\
&& {\bf A}(n_1,...,n_i,...,n_d|n_1,...,n_i-1,...,n_d)
   =\left( \begin{array}{cc} {\bf 0} & {\bf 0}  \\
   -{\bf U}_i & {\bf 0}  \end{array} \right)  \nonumber \\
&& {\bf A}(m_1,...,m_d|n_1,...,n_d)= \left( \begin{array}{cc}
   {\bf 0} & {\bf 0} \\ {\bf 0} & {\bf 0} \end{array} \right) ,
   \hskip .5cm {\rm otherwise}
\end{eqnarray}
where $\bf 0$ is  the $d$-dimensional  zero matrix, $\bf W$ a $d\times d$
matrix with elements ${W}_{ij}=\omega_{i,j+d}$,
and ${\bf U}_i$ a $d\times d$ matrix whose only nonzero
element is ${\lbrace U_i \rbrace }_{i i}=1$.

With periodic boundary conditions, matrices in (7) are invariant
when indices $m_i, n_i$ are changed by $N_i$. Then the matrix
is block-diagonal in the Fourier space.  This leads to the
following expression for the partition function
\begin{eqnarray}
Z && = \sqrt {{\rm det} {\bf A}}  \nonumber \\
  && =\prod_{n_1=1}^{N_1}\cdots
     \prod_{n_d=1}^{N_d}\biggl[{\rm det} \left( \begin{array}{cc}
     {\bf 0} & {\bf B}  \\  -{\bf B}^* & {\bf 0}
     \end{array}  \right) \biggr]^{1/2} \nonumber  \\
  && =\prod_{n_1=1}^{N_1}\cdots
     \prod_{n_d=1}^{N_d}{\rm det} \bigl| {\bf B}\bigr|,
\end{eqnarray}
where ${\bf B}\equiv{\bf B}(2\pi n_1/ N_1, ..., 2\pi n_d/N_d)$
is a $d\times d$ matrix with elements
\begin{equation}
\begin{array}{rll}
B_{ij}
& = \omega_{i,j+d},  & \mbox{$i\not= j$}  \\
& = \omega_{i,i+d}+e^{2\pi i n_i/N_i}, \>\>
	       & \mbox{$i=j  , \hskip 1cm i,j=1,...,d,$}
\end{array}
\end{equation}
and ${\bf B}^*$ is the complex conjugate of $\bf B$.
In this way, we obtain the per-site free energy
\begin{eqnarray}
f_d && =\lim_{N_i\rightarrow \infty} N^{-1} \ln Z \nonumber \\
    && ={1\over {(2\pi)^d}}\int^{2\pi}_0 d\theta_1 \cdots
	\int^{2\pi}_0 d\theta_d
	\ln \big|D_d(\theta_1,\cdots,\theta_d)\big|,
\end{eqnarray}
where
\begin{equation}
D_d (\theta_1,\cdots,\theta_d)
= {\rm det}\{{\bf B}(\theta_1,\cdots,\theta_d)\},
\end{equation}
Particularly, for $d=2$
and $3$, we have
\begin{eqnarray}
D_2(\theta_1, \theta_2)
& = & \omega_{1234}+\omega_{24}e^{i\theta_1} +\omega_{13}e^{i\theta_2}
   +e^{i(\theta_1+\theta_2)}, \nonumber \\
D_3(\theta_1, \theta_2, \theta_3)
& = & \omega_{123456}+\omega_{2356}e^{i\theta_1} +\omega_{1346}e^{i\theta_2}
   +\omega_{1245}e^{i\theta_3} \nonumber \\
& \  & +\omega_{14}e^{i(\theta_2+\theta_3)}+
   \omega_{25}e^{i(\theta_1+\theta_3)}+ \omega_{36}e^{i(\theta_1+\theta_2)}
   +e^{i(\theta_1+\theta_2+\theta_3)}.
\end{eqnarray}
Here, the $\omega$'s with 4 or more indices are those defined by (2) - (4b).
Particularly, we have,
in addition to those given in (4a) and (4b),
$\omega_{2356} =\omega_{25}\omega_{36} -\omega_{26}\omega_{35},
\omega_{1346} =\omega_{14}\omega_{36} -\omega_{16}\omega_{34}.$
Explicitly,
$D_d(\theta_1,\cdots,\theta_d)$
is a linear
combination of terms in the form of
$c_{ij\cdots k}e^{i(\theta_i+\theta_j\cdots+\theta_k)}$,
where
the coefficient $c_{ij\cdots k}$ is the determinant of the matrix
${\bf W}$
with the $i, j, ...,k$th rows
and columns deleted. By taking $\omega_{ij} = \sqrt {z_iz_j}, \>
z_i=z_{i+d}$ for which
the free-fermion weights with 4 or more indices
vanish identically, these results reduce
to those of  \cite{rf8}.

\section{THE CRITICAL BEHAVIOR}

For $d=2$ the vertices are those of a six-vertex model shown
in Fig. 4 with the weights
\begin{equation}
\begin{array}{ll}
\Omega_1=1, \hskip 1.6cm & \Omega_2=\omega_{1234} =
		   \Omega_3\Omega_4-\Omega_5\Omega_6  \\
\Omega_3=\omega_{13}, & \Omega_4 = \omega_{24} \\
\Omega_5=\omega_{14}, & \Omega_6 = \omega_{23}.
\end{array}
\end{equation}
Note that the free-fermion condition is now
$\Omega_1\Omega_2 = \Omega_3\Omega_4-\Omega_5\Omega_6$
and differs slightly (corresponding to the negation
of $\Omega_2$) from the usual form \cite{rf9} $\Omega_1\Omega_2 =
- \Omega_3\Omega_4+\Omega_5\Omega_6$
for which the partition function is defined by (5)  without the
loop factor $(-1)^\ell$.  Rewriting (12) and carrying out one
integration, say, over $\theta_2$,  we have
\begin{eqnarray}
f_2 && = {1\over {(2\pi)^2}}
	 \int^{2\pi}_0 d\theta_1\int^{2\pi}_0 d\theta_2
	 \ln \big| \Omega_2+\Omega_3e^{i\theta_1} +\Omega_4e^{i\theta_2}
	 +\Omega_1e^{i(\theta_1+\theta_2)} \big| \nonumber \\
    && =  {1\over {2\pi}}\int^{2\pi}_0 d\theta_1
	 \ln {\rm max} \biggl\{\bigl| \Omega_2+\Omega_3e^{i\theta_1}\bigr|,
	 \> \bigl|\Omega_4 +\Omega_1 e^{i\theta_1} \bigr| \biggr\}.
\end{eqnarray}
This free energy reduces to that of usual free-fermion
6-vertex model after negating $\Omega_2$ \cite{rf8}.
The equivalence of the two free energies (with and without the loop
factor) is a property unique to two dimensions.

When one of the two absolute values in (14) is larger than the other
for all $\theta_1$,
then one can  carry out the remaining integration and
obtains
\begin{equation}
f_2= {\rm max} \{ |\Omega_1|, |\Omega_2|,|\Omega_3|, |\Omega_4|\},
\end{equation}
so that the system is in a "frozen" state.
Now the two absolute values in (14) are linear in $\cos \theta_1$.
As a consequence, when perturbed from the frozen states (15),
the two absolute values will cross
only at $\theta_1 =0$ or $\pi$, near which one of the two
absolute values is always larger.
This leads to the two critical conditions
\begin{equation}
\bigl|\Omega_2 \pm\Omega_3\bigr| =\bigl|\Omega_4\pm \Omega_1\bigr| ,
\end{equation}
at which $f_2$ becomes singular.
Near the critical point say, $\Omega_2+\Omega_3=\Omega_4+\Omega_1$ for all
$\Omega_i>0$, for instance, let
\begin{equation}
t=\Omega_2 + \Omega_3 -\Omega_4- \Omega_1\sim {\rm small\>\>positive}.
\end{equation}
Then, the singular part of $f_2$ is given by
\begin{equation}
\{f_2\}_{\rm sing} = {1\over {2\pi}}
\int^{\alpha(t)}_{-\alpha(t)} d\theta_1
\ln\biggl| {{\Omega_2+\Omega_3e^{i\theta_1}}
\over {\Omega_4+\Omega_1e^{i\theta_1}}}\biggr|,
\end{equation}
where $\alpha(t)>0$ is determined from
\begin{equation}
\bigl| \Omega_2+\Omega_3e^{i\alpha}\bigr|
= \bigl| \Omega_4+\Omega_1e^{i\alpha}\bigr|.
\end{equation}
It is easy to verify that $\alpha \sim \sqrt {t}$,
and for small $t$ the integrand in (18) is of the order of
$t+\theta_1^2$.  It follow that
\begin{equation}
\{f_2\}_{\rm sing} \sim t^{3/2}.
\end{equation}
The same critical behavior is deduced  if any of the weights,
say, $\Omega_2$, is negative.

The example of $d=2$ serves to illustrate the origin of
the critical behavior and the critical point.
After carrying out one-fold integration, $f_2$ is given in the form of
an integration of a competition of the logarithms of
 two absolute values as in the second line of
(14).  Then, the critical point is the point
at which the two absolute values
are equal and the integrand in (14) switches from one
absolute value to the other for small deviations from the
critical point.  This creates a singularity in the free energy.
We shall call this the switching property of the integrand.
A prerequisite for switching to occur  is that the ratio of the
two absolute values
is an extremum in $\theta_1$ at the critical point.
For $d=2$ this can occur only at $\theta_1=0$ or $\pi$.
This leads to a singular part of the free energy in the form of (18),
where the integration is taken over a region ${\cal R}$
whose boundary is determined by setting the two absolute  values
equal at fixed small deviation $t$
from criticality.
It follows that the
singular behavior is  that of this
integral, namely,
$tV(t)$, where $V(t) \sim \sqrt t$ (for $d=2$)
is the volume
of ${\cal R}$.

Applying the same  argument
to $d\geq  3$, one first carries out one-fold
integration
over, say, $\theta_j$,
yielding $f_d$ in the form of an integration of a competition
of the logarithms of the absolute values of
 $A_j(\theta_\alpha),\>\alpha\not= j$, the collection of
 terms in $D_d$ linear in $e^{i\theta_j}$,
and $B_j(\theta_\alpha)$,
the collection of terms in $D_d$ independent of $e^{i\theta_j}$.
Then $f_d$ can be singular at points in the parameter space at which the
two absolute values become equal and possess the aforementioned
switching and extremum property.  Namely,
there exist angles $\theta_{\alpha 0}, \>\alpha \not=j$, such that
\begin{equation}
\bigl|A_j(\theta_{\alpha 0})\bigr| =
 \bigl|B_j(\theta_{\alpha 0})\bigr| ,\hskip 1cm t=0
\end{equation}
and
\begin{eqnarray}
 && \bigl|A_j(\theta_{\alpha })\bigr| >
 \bigl|B_j(\theta_{\alpha })\bigr| , \hskip 1cm t<0 \nonumber \\
 && \bigl|A_j(\theta_{\alpha })\bigr| \leq
 \bigl|B_j(\theta_{\alpha })\bigr| , \hskip 1cm t=0+
\end{eqnarray}
for  $\theta_\alpha$ in a neighborhood of $\theta_{\alpha 0}$,
where $t$ is the deviation from the critical point.
(Here, the roles of $A_j$ and $B_j$ can be interchanged.)
Since the integrand of the free energy switches from $|A_j|$ to $|B_j|$
for $t=0+$,
a singular part in the form of
(18) appears in the free energy,
which is the integration of $\ln |A_j/B_j|$
over a region ${\cal R}$ bounded by
\begin{equation}
\bigl|A_j(\theta_{\alpha })\bigr| =
 \bigl|B_j(\theta_{\alpha })\bigr| ,\hskip 1cm t={\rm small
\>\>fixed}.
\end{equation}
The critical condition
is  now given by (21) and the critical behavior is
$tV(t)$, where $V(t)$
is the volume of the $(d-1)$-dimensional region ${\cal R}$.

To determine $V(t)$, we expand
$\ln |A_j/B_j|$
about $\theta_{\alpha 0}$ and $t=0$,
obtaining  generally the leading behavior
\begin{equation}
ct+F(\Delta\theta_\alpha),
\end{equation}
where $c$ is a constant and
$\Delta \theta_\alpha = \theta _\alpha-\theta_{\alpha 0}$.
Here, due to the extremum property,
$F$ is a quadratic form   in variables $\Delta\theta_\alpha$,
which is either positive or negative
definite.
This implies that the
region ${\cal R}$ has linear dimensions $\Delta\theta_\alpha
\sim t^{1/2}$, and hence $V(t) \sim t^{(d-1)/2}$. The singular
part of the free energy  behaves as
\begin{equation}
\{f_d\}_{\rm sing} \sim t^{(d+1)/2}.
\end{equation}
This is
the critical behavior found in the
model $\Omega_2 =0$ \cite{rf8}.
The meaning of the singular behavior (25) for $d={\rm odd}$
is that the $[(d+1)/2]$-th
derivative of $f_d$ is discontinuous.

However, the singular behavior (25) may not be the dominate one.
If the $(d-1)$-dimensional
integral $\int \ln |A_j|$
dominant in (21) - (23) is also singular at $t=0$,
then the critical behavior of the free energy at $t=0$ is
that of a
$(d-1)$-dimensional model.
By the same token,
the critical behavior can be reduced further to that of lower dimensions.
Examples are given in the next section.

\section{EXAMPLES}

\subsection{Example 1} If lines in the vertex model are nonintersecting,
then we have the
model considered in \cite{rf8} with
\begin{equation}
D_d (\theta_1, \cdots, \theta_d) = 1+\sum_{i=1}^d z_i e^{i\theta_i}.
\end{equation}
There is no loss of generality in supposing $|z_i|\leq 1$.
Carry out the integration in
(10) over $\theta_1$, say, after rewriting
$D_d$
as $e^{-i\theta_1}+z_1 +z_2e^{i\theta_2} +\cdots +z_de^{i\theta_d} $, where
we have renamed $\theta_\alpha -\theta_1$ as $\theta_\alpha$.
If any of the $z_\alpha$ is negative,
one replaces $\theta_\alpha$  by $\pi + \theta_\alpha$ and obtains
$A_1=1$ and $B_1=|z_1| +|z_2|e^{i\theta_2}  +\cdots +|z_d|e^{i\theta_d}$.
It is then straightforward to verify that the extremum property occurs only
at $\theta_{\alpha 0} =0,\>\alpha=2,..., d.$
The critical point occurs at
\begin{equation}
| z_1| + |z_2| + \cdots +|z_d|  =1,
\end{equation}
and the critical behavior of the free energy is $t^{(d+1)/2}$.

\subsection{Example 2}. Consider the $3$-dimensional model with weights
\begin{eqnarray}
&&  \omega_{36}=z, \hskip 1.5cm \omega_{24}=z_1/z \nonumber \\
&&  \omega_{16}=\omega_{26}=z \nonumber \\
&&  \omega_{34}=\omega_{35}=1 \nonumber \\
&&  \omega_{14}=\omega_{25}=\omega_{15}=0
\end{eqnarray}
for  $z$ and $z_1$ real,  implying, using (4a) and (4b),
\begin{eqnarray}
&&  \omega_{123456}=z_1, \hskip 0.5cm \omega_{1245}=0 \nonumber \\
&&  \omega_{1346}=\omega_{2356}=-z.
\end{eqnarray}
By appropriately changing integration variables, the per-site free
energy (10) assumes the form
\begin{equation}
f_3 = {1 \over {{(2\pi)}^3}} \int^{2\pi}_0 d\theta_1
	  \int^{2\pi}_0 d\theta_2  \int^{2\pi}_0 d\theta_3
	  \ln \big| z_1 +ze^{i\theta_1}+ze^{i\theta_2}
	  +ze^{i(\theta_1+\theta_2)}
	  +e^{i(\theta_1+\theta_2+\theta_3)} \big| .
\end{equation}
After carrying out the integration over $\theta_3$, one obtains
$A_3=1$, $B_3=z_1+ze^{i\theta_1}+ze^{i\theta_2}
+ze^{i(\theta_1+\theta_2)}$ with
\begin{eqnarray}
{\bigl|B_3\bigr|}^2 & = & (z_1+3z)^2
       +2z(z+z_1)(\cos\theta_1+\cos\theta_2 -2) \nonumber \\
&  & \mbox{} +2zz_1\bigl[\cos(\theta_1+\theta_2) -1\bigr]
       +2z^2 \bigl[\cos(\theta_1-\theta_2)-1\bigr].
\end{eqnarray}
It can be verified that  ${|B_3|}$ possesses one extremum at $\theta_{10}
=\theta_{20}=0$ in the regimes $z_1/z>-1/3$ and $z_1/z<-3$,
which, by setting the extremum equal to one, leads to the critical point
\begin{equation}
\big|z_1+3z \big| =1.
\end{equation}
Define $t=|z_1+3z|-1$.  Then one finds
\begin{equation}
\begin{array}{ll}
f_3                 =0,                    & t<0  \\
\{f_3\}_{\rm sing}  \sim t^2, \hskip 1.9cm & t=0+
\end{array}
\end{equation}
for $z_1/z>-1/3$,
and
\begin{equation}
\begin{array}{ll}
f_3     =\ln |z_1|,                          & t>0 \nonumber \\
\{f_3\}_{\rm sing} \sim {t}^2, \hskip 1.9cm  & t=0-.
\end{array}
\end{equation}
for $z_1/z<-1/3$.
In addition, another extremum occurs  at
$\cos \theta_{10}=\cos \theta_{20}= - (z_1+z)/2z_1 $ in the regime
 $|1-3z_1/z|>2$,
which leads to the critical point
\begin{equation}
{(z_1-z)}^3 =z_1.
\end{equation}
Define $t={|z_1-z|}^3-|z_1|$. Then one finds the behavior $(33)$ for
$z_1/z<-1/3$ and $(34)$ for $z_1/z>-1/3$.
These results lead to the phase diagram shown in Fig. $5$.

\subsection{Example 3}
Consider another 3-dimensional model with weights
\begin{eqnarray}
&& \omega_{15}=z, \hskip 2.6cm \omega_{36}=-1 \nonumber \\
&& \omega_{14}=\omega_{25}=0 \nonumber \\
&& \omega_{24}=\omega_{26}=\omega_{34}=\omega_{35}=\omega_{16}=1
\end{eqnarray}
implying,
\begin{eqnarray}
&& \omega_{2356}=\omega_{1346}=-1 \nonumber \\
&& \omega_{1245}=-z \nonumber \\
&& \omega_{123456}=2z+1.
\end{eqnarray}
The per-site free energy is
\begin{equation}
f_3={1 \over {(2\pi)}^3} \int_0^{2\pi} d\theta_1
			   \int_0^{2\pi} d\theta_2
			   \int_0^{2\pi} d\theta_3
       \ln \big| 2z+1-e^{i(\theta_1+\theta_2)}- e^{i\theta_1}
	   - e^{i\theta_2} -z e^{i\theta_3}
	   +e^{i(\theta_1+\theta_2+\theta_3)} \big|
\end{equation}
After carrying out the integration over $\theta_3$, one obtains
$A_3=z-e^{i(\theta_1+\theta_2)}$,
$B_3=2z+1-e^{i(\theta_1+\theta_2)}-e^{i\theta_1}-e^{i\theta_2}$,
 leading to the
critical point
\begin{equation}
z=1,
\end{equation}
at $\theta_{10}=\theta_{20}=0$.
Define $t=z-1$. It can be verify that $|A_3| \leq |B_3|$ for $t\geq 0$,
and a switching from $|B_3| $ to  $|A_3|$ occurs in
a small regime in the neighborhood of $\theta_{10}=-\theta_{20}$,
$t=0-$. Hence the free energy contains a singular part of the order of
$O(t^2)$ for $t=0-$.  However, it turns out that the dominate
integral
\begin{equation}
f_3\sim  {1 \over {(2\pi)}^2} \int_0^{2\pi} d\theta_1
\int_0^{2\pi} d\theta_2 \ln |B_3(\theta_1, \theta_2)|,
\end{equation}
is also singular at $t=0$.
As in (14), further integrations of (40)
lead to the
critical behavior
\begin{eqnarray}
f_3  && = \ln |2z+1|, \hskip 5.3cm t\geq 0 \nonumber \\
     && = \ln |2z+1|+
	  {{2\sqrt 2}\over {3\pi}}\bigl| t\bigr|^{3/2},
	  \hskip 3cm t=0-.
\end{eqnarray}
This critical behavior is verified
by numerical integrations whose results are shown
in Fig. 6.

\section{SUMMARY}

We have solved exactly a free-fermion vertex model in $d$
dimensions.  The vertex model is described by lines running
in a preferred direction with the weights of vertices of
intersecting lines prescribed by the free-fermion condition (2).
The partition function (5)   contains
a fugacity $-1$ for each loop
of lines and therefore consists of positive and negative
weights.  The per-site free energy is  evaluated  exactly in (10).
The critical point is found to be given by (21) and the critical
behavior given by $t^{(d+1)/2}$.  In exceptional cases this critical behavior
is modified with   $d$
replaced by an integer lower than the actual dimensionality.

This work has been supported in part by NSF Grants DMR-9313648 and
INT-9207261.

\begin{figure}
\caption{Decomposition of a vertex configuration of
	 $2r$ incident bonds into $r!$ line configurations.}
\label{fig1}
\end{figure}

\begin{figure}
\caption{A city of $2d$ points for $d=3$.}
\label{fig2}
\end{figure}

\begin{figure}
\caption{The correspondence between vertex configurations on ${\cal L}$
	 and dimer coverings on ${\cal L}'$.}
\label{fig3}
\end{figure}

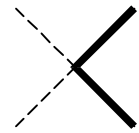
\begin{figure}
\caption{Vertex configurations and weights for a 6-vertex model.}
\label{fig4}
\end{figure}

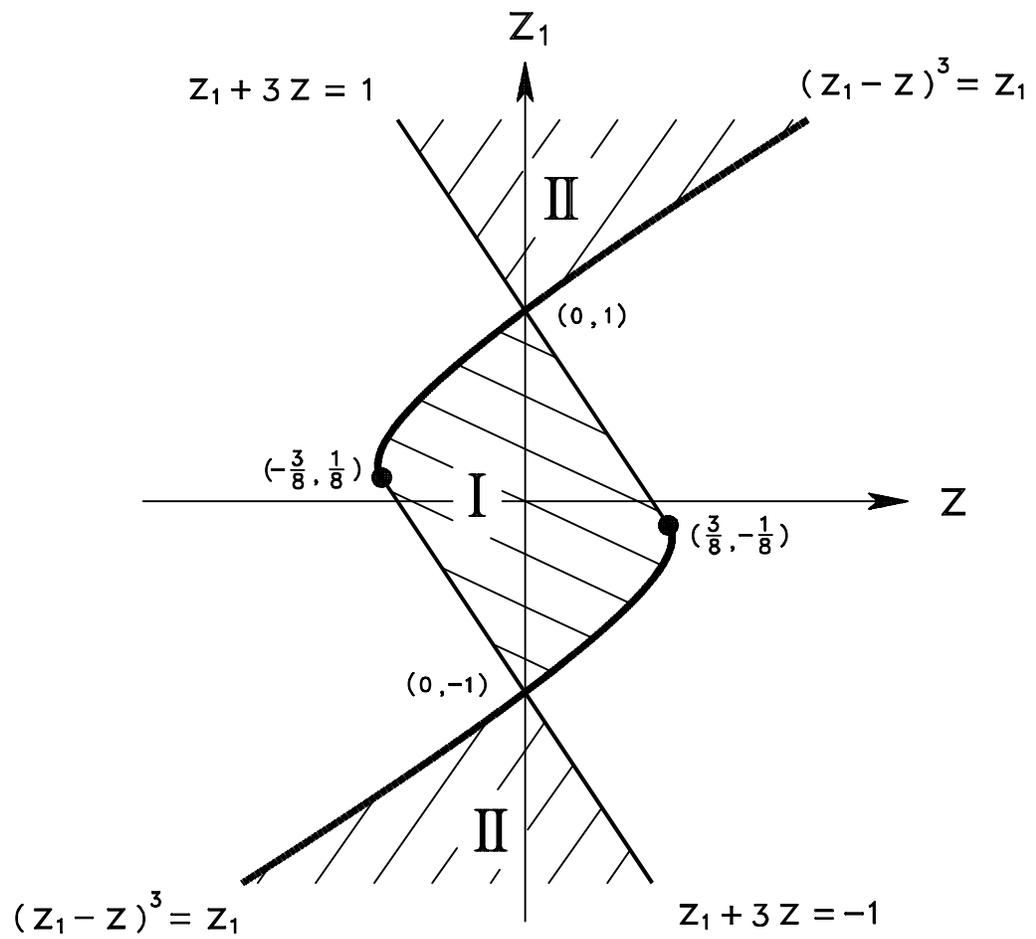
\begin{figure}
\caption{Phase diagram of the three-dimensional model (30).
	 Regimes I and II are in frozen states with $f_3=0$ and $f_3=
	 \ln |z_1| $, respectively.}

\label{fig5}
\end{figure}

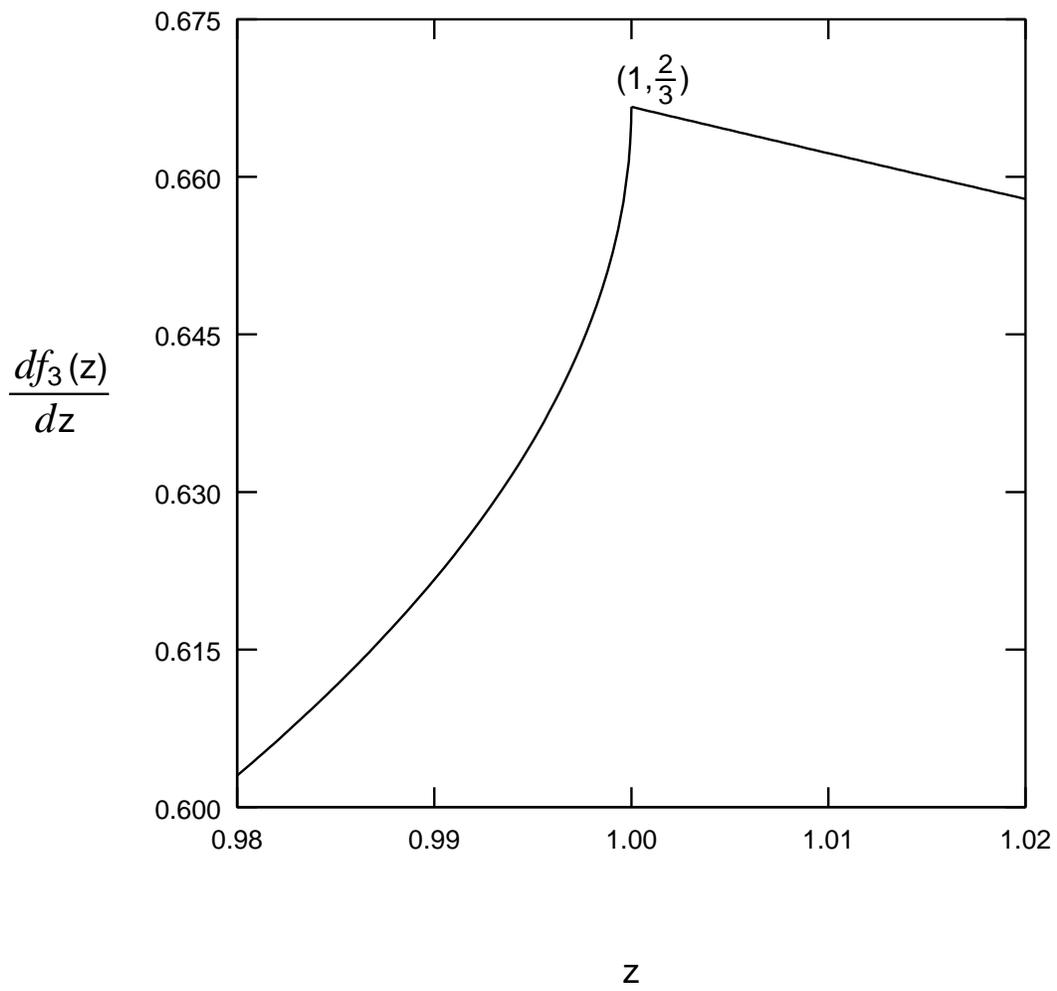
\begin{figure}
\caption{Results of  numerical integrations of
	 $df_3(z)/dz$
	 for the three-dimensional model (38).}

\label{fig6}
\end{figure}

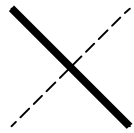
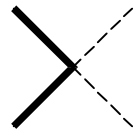
\end{document}